\documentclass[11pt]{article}
\usepackage{fleqn,cospar}

\usepackage{url}

\usepackage{graphicx}
\usepackage[figuresright]{rotating}

\title{Magnetohydrodynamic Simulations of Accretion Disks around a Weakly Magnetized Neutron Star in Strong Gravity}

\author{Y. Kato\address{Graduate School of Science and Technology,
Chiba University, \\1-33 Yayoi-cho, Inage-ku, Chiba 263-8522, Japan},
        M. R. Hayashi\address{National Astronomical Observatory,
	Mitaka, Tokyo 181, Japan},
        S. Miyaji$^{1}$,
        and
        R. Matsumoto$^{1}$} 

\begin{document}

% typeset front matter
\maketitle

\begin{abstract}
We carried out two dimensional high-resolution magnetohydrodynamic
(MHD) simulations of an accretion disk around a weakly magnetized
neutron star. General relativistic effects are taken into account by
using pseudo-Newtonian potential of Pacz\'{y}nski, B., P. J. Witta
(1980).  When magnetic loops connect the neutron star and the
accretion disk, the twist injection from the disk or from the rotating
neutron star triggers expansion of the loops.  Since the expanding
magnetic loops prevent inflow toward the magnetic poles of the neutron
star, disk matter accumulates on the boundary between the
magnetosphere and the disk.  Magnetic reconnection taking place in the
loops creates a channel along which the disk matter can accrete and
unloads the magnetosphere.  This process produces quasi-periodic
variation of the accretion flow in the innermost region of the disk.
We found two kinds of oscillations.  One is the magnetospheric
oscillation regulated by magnetic reconnection. The other is the
radial disk oscillation. The typical frequency of the oscillations is
100 Hz to 2 kHz.  Furthermore, we predict that QPO sources inevitably
accompany X-ray flares by magnetic reconnection and bipolar outflows
of hot X-ray emitting plasma similar to the optical jets in
protostars.
\end{abstract}

\section*{INTRODUCTION}

Shortly after the quasi-periodic oscillations of low mass X-ray
binaries were found, various models have been proposed (e.g., Lamb, F.
K., N. Shibazaki, M., M. A. Alpar, and J. Shaham, 1985;  Miller, M. C.,
F. K. Lamb, and D. Psaltis, 1998).  Although the disk-magnetosphere
interaction is believed to be essential in such models, its detailed
physical mechanisms have not been worked out yet.

Nonlinear simulations of the interaction between the dipole magnetic
field of a star and its surrounding disk were carried out by Hayashi,
M. R., K. Shibata, and R. Matsumoto (1996).   By assuming Newtonian
gravity and neglecting the stellar rotation, they showed that the
magnetic interaction can explain the X-ray flares and outflows
observed in protostars. When the dipole magnetic field of the star is
twisted by the rotation of the disk, the magnetic loops connecting the
star and the disk expand. Magnetic reconnection taking place in the
expanding loops produces X-ray flares and hot plasma outflows.  We
extend this model to the neutron star.  The dynamics around the
marginally stable radius in the strong gravity of the neutron star is
mimicked by using pseudo-Newtonian potential of the form:
$\Psi_{PN}=-GM/(R-r_{g})$ where $r_{g}(=2GM/c^{2})$ is the
Schwartzschild radius and $R(=\sqrt{r^{2}+z^{2}})$ the distance from
the center of the neutron star.  In this paper, we primarily concern
with the magnetic interaction around the boundary between the
magnetosphere and the accretion disk surrounding the neutron star. 

\section*{SIMULATION MODELS}

We numerically solved axisymmetric MHD equations in a cylindrical
coordinate by using a 2-D MHD code based on a modified Lax-Wendroff
scheme with artificial viscosity.

\begin{figure}
\vspace{-8mm}
\begin{minipage}{90mm}
\begin{tabular}{ccc}
\hline
\\
Model Parameters & Model-A & Model-B \\ \hline
NS mass ($M_{\odot}$) 		& \multicolumn{2}{c}{1.4} \\
NS radius ($r_{g}$)		& \multicolumn{2}{c}{2.0} \\
Corotation radius ($r_{g}$)	& 13  & $\infty$ \\
Plasma $\beta$                  & 10  & 100 \\
Magnetic Reynolds Number	& \multicolumn{2}{c}{$R_{m}=1000$} \\
Magnetic Diffusivity		& \multicolumn{2}{c}{$\eta = 1/R_{m}$} \\
Center of torus ($r_{g}$)	& \multicolumn{2}{c}{13}  \\
Computational area ($r_{g}$)    & \multicolumn{2}{c}{$30\times 30$} \\
Number of Grids 		& \multicolumn{2}{c}{$500\times 500$} \\
\hline
\end{tabular}
  \label{model_param:table}
  \vspace{1cm}
  {\sf Table. 1. Model Parameters}
\end{minipage}
\hfil\hspace{\fill}
\begin{minipage}{75mm}
  \includegraphics[width=75mm]{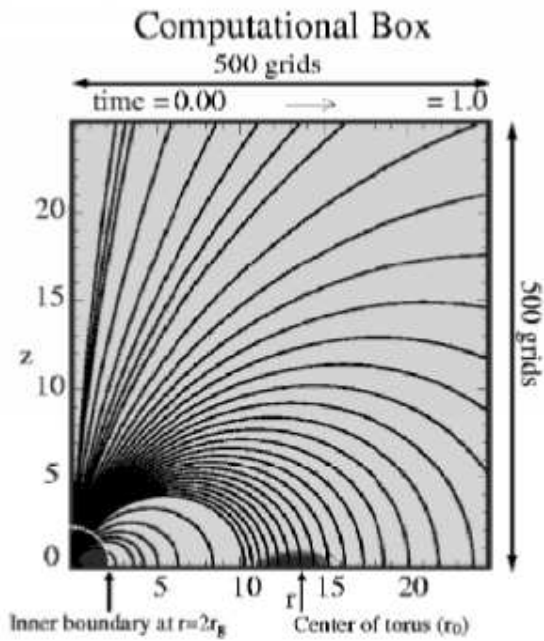}
  \caption{Initial State of Numerical Simulation}
  \label{initial:ps}
\end{minipage}
\end{figure}

Assuming that a neutron star is an aligned rotator, the dipole magnetic
moment of the neutron star is aligned with stellar rotation axis.  We
put a cold sub-Keplerian disk around the star (see Figure
\ref{initial:ps}).  We adopt polytropic equation of state
$P=K\rho^{5/3}$ for the disk.  The radius of the density maximum of
the disk is 13 $r_{g}$.  The inner edge of the disk initially locates
at 11 $r_{g}$.  Outside the disk, we assumed non-rotating, spherical,
and isothermal hot halo.  The radius of the neutron star is assumed to
be smaller than the marginally stable radius.  We impose the absorbing
boundary condition on the surface of a neutron star at $R=2.0 r_{g}$.
The strength of the dipole magnetic moment is parameterized by plasma
$\beta$ defined as the ratio of gas pressure to magnetic pressure at
$(r,z)=(13 r_{g},0)$.  The unit of length is Schwartzschild radius,
and the velocity is normalized by the speed of light [$r_{g}=c \hbox{
(the speed of light)}=1$].  We divided the computational area into
$500\times 500$ grids and solved axisymmetric resistive MHD equations
by applying a 2-D resistive MHD code which is originally developed by
Hayashi, M. R., K. Shibata, and R. Matsumoto (1996). Resistivity is
assumed to be uniform. 

\section*{SIMULATION RESULTS}

We carried out simulations of the following two models.  Model-A is
for a rotating neutron star and model-B is for a non-rotating neutron
star.  These model parameters are summarized in table 1.  The reason
why we choose these two models is that the angular momentum transport
between the neutron star and the accretion disk intrinsically depends
on the differential rotation between them.

\subsection*{Model-A : Fast-Rotator}

\begin{figure}[t]
\begin{minipage}{180mm}
\begin{minipage}{50mm}
  \includegraphics[width=50mm]{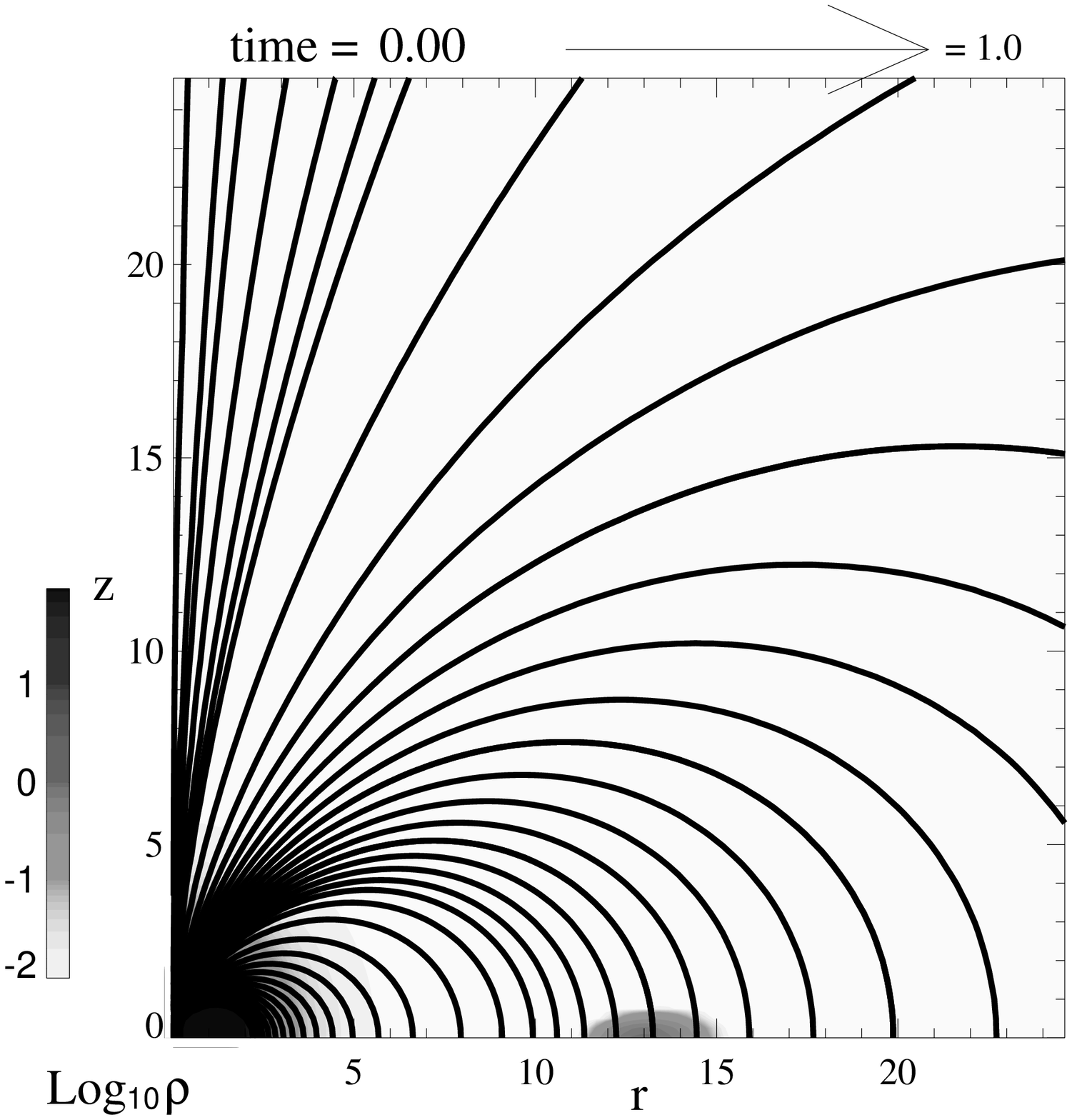}
\end{minipage}
\hfil\hspace{\fill}
\begin{minipage}{50mm}
  \includegraphics[width=50mm]{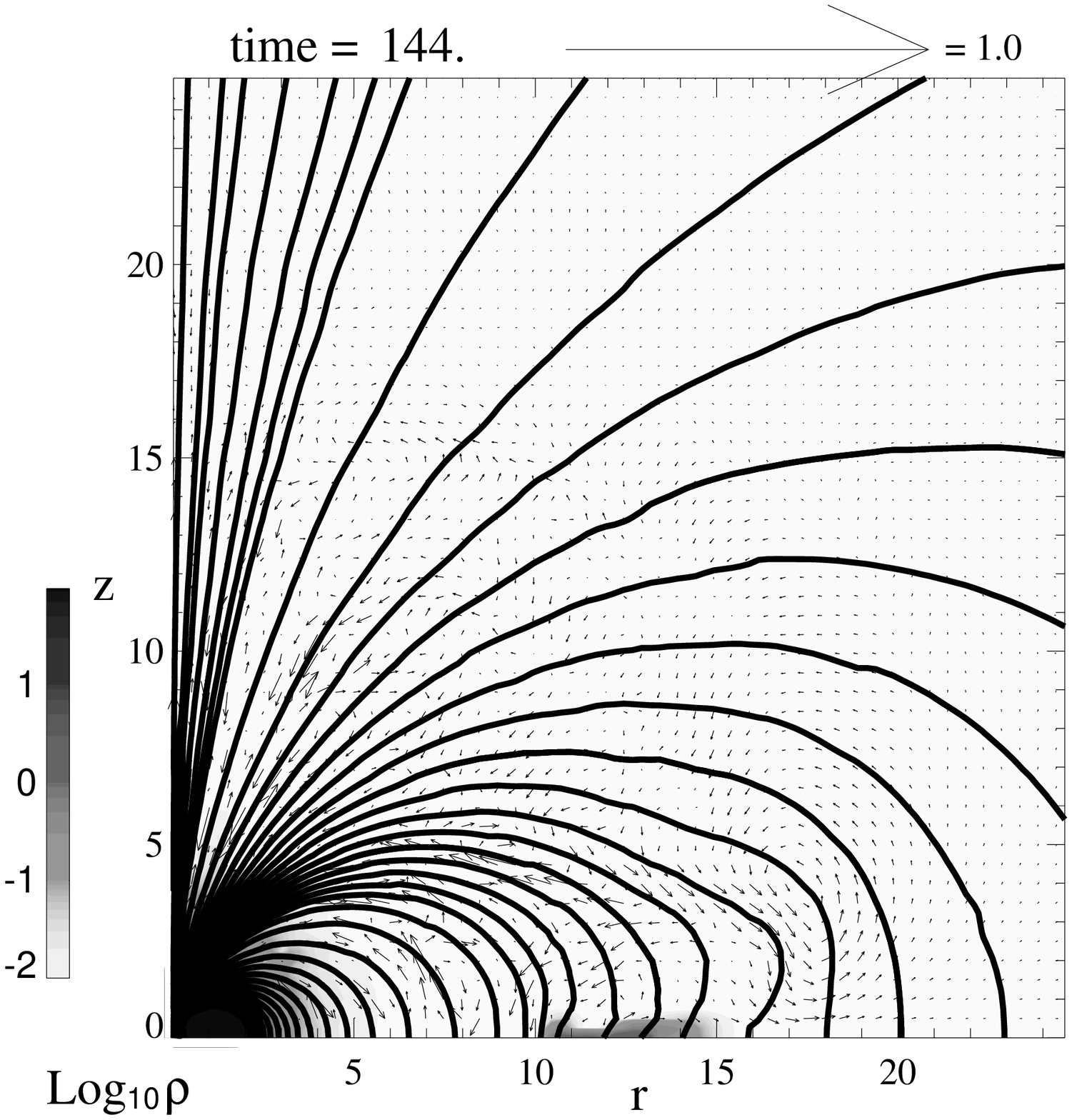}
\end{minipage}
\hfil\hspace{\fill}
\begin{minipage}{50mm}
  \includegraphics[width=50mm]{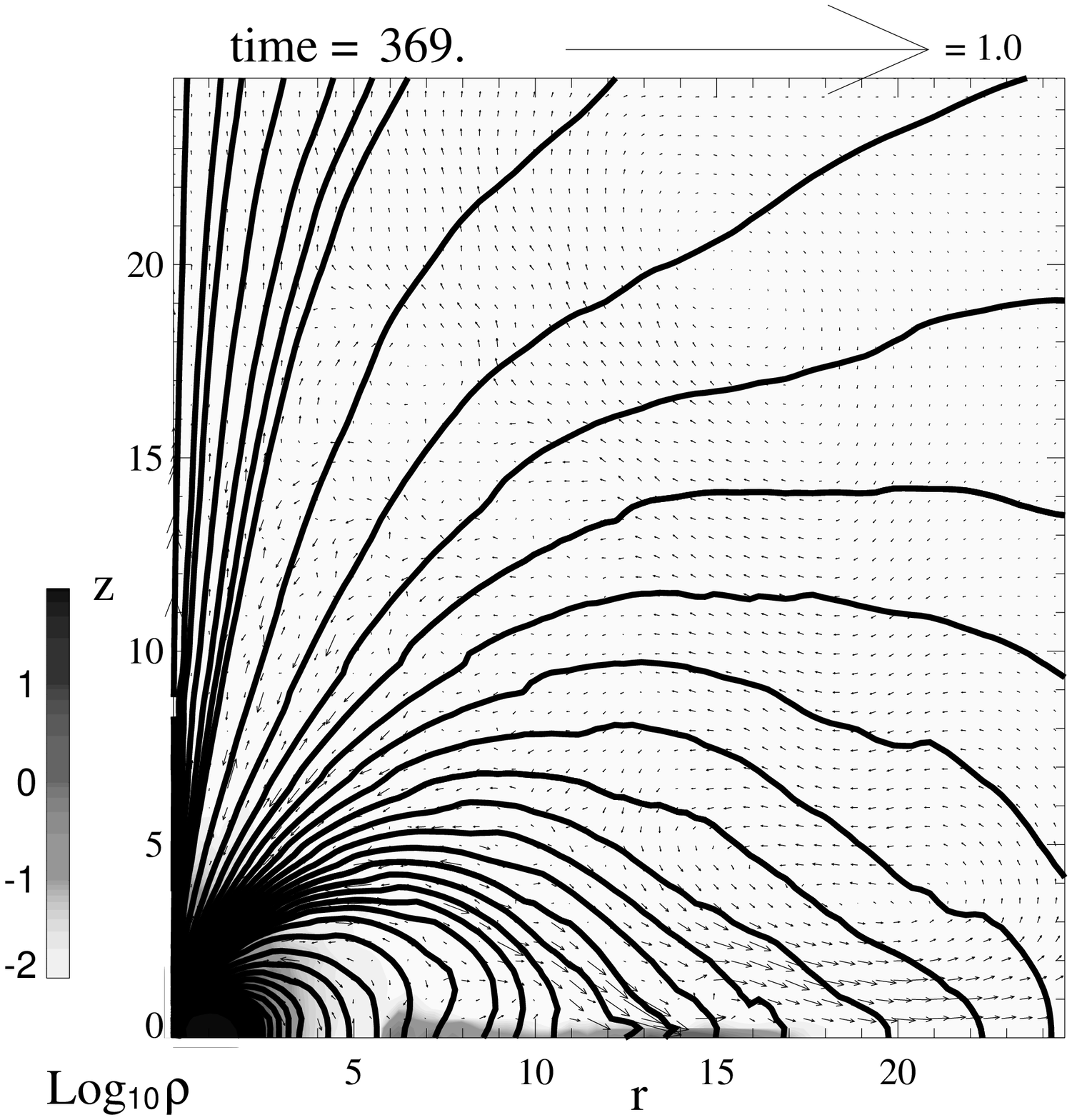}
\end{minipage}
\hfil\hspace{\fill}
\begin{minipage}{50mm}
  \includegraphics[width=50mm]{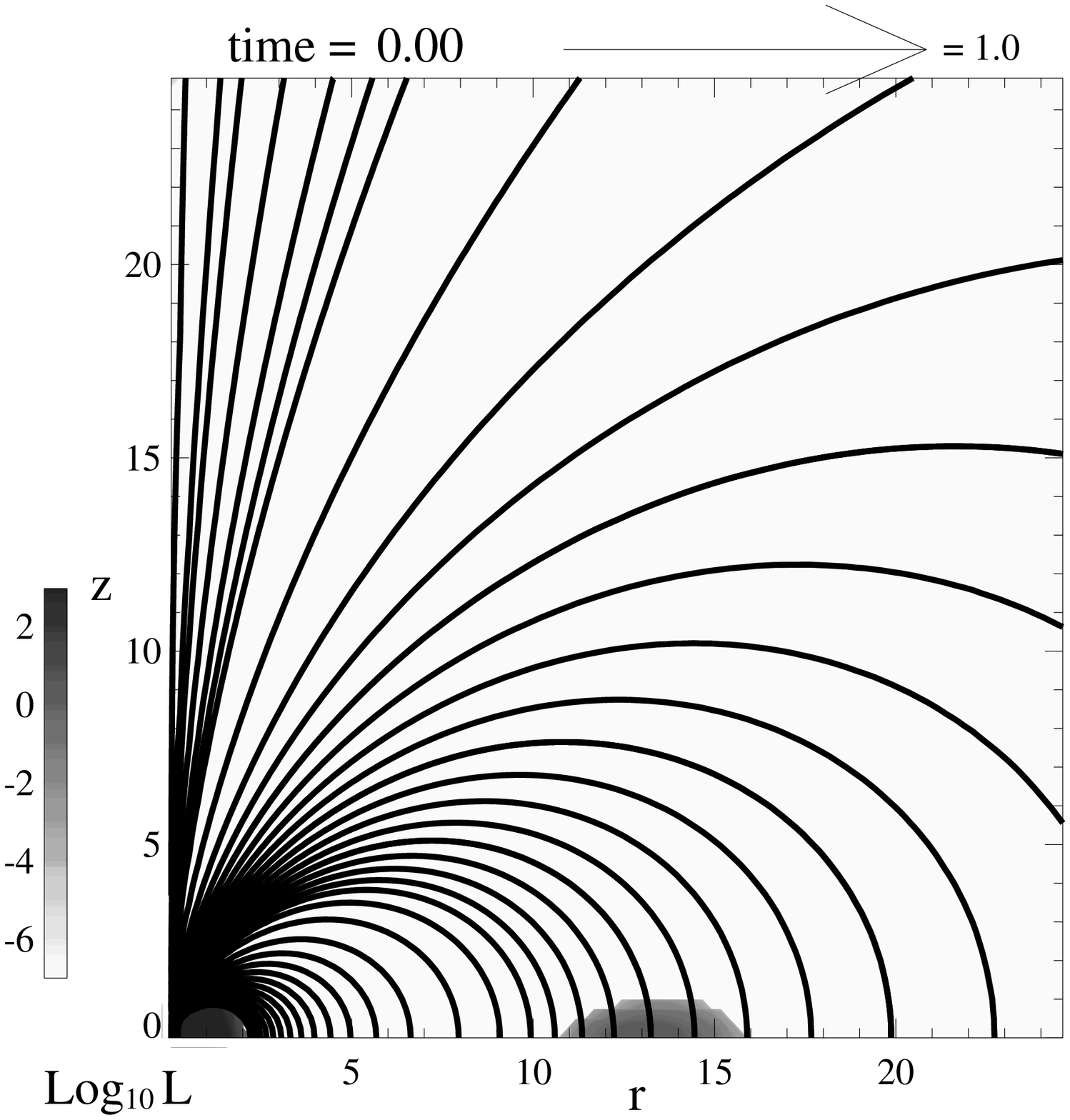}
\end{minipage}
\hfil\hspace{\fill}
\begin{minipage}{50mm}
  \includegraphics[width=50mm]{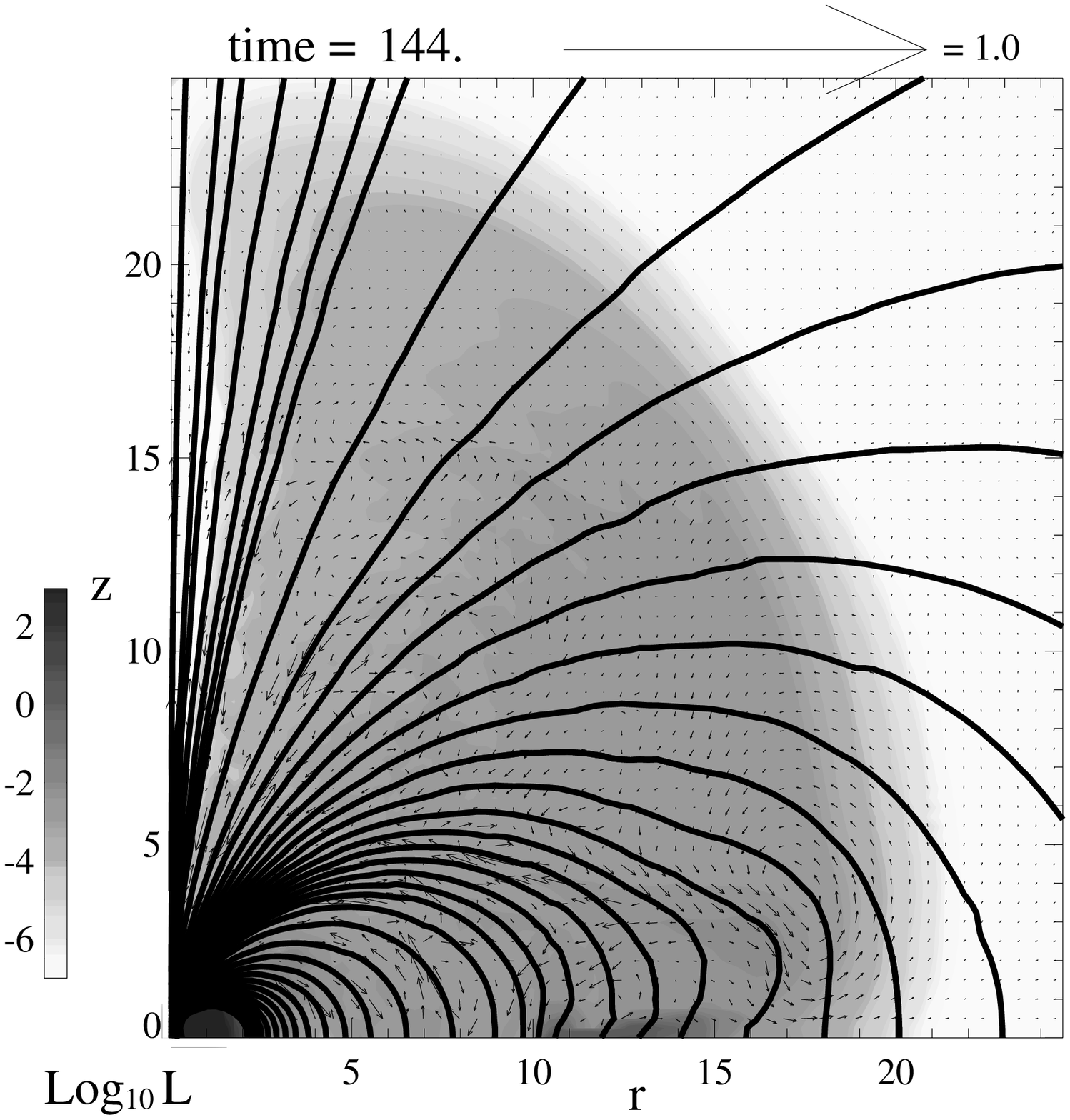}
\end{minipage}
\hfil\hspace{\fill}
\begin{minipage}{50mm}
  \includegraphics[width=50mm]{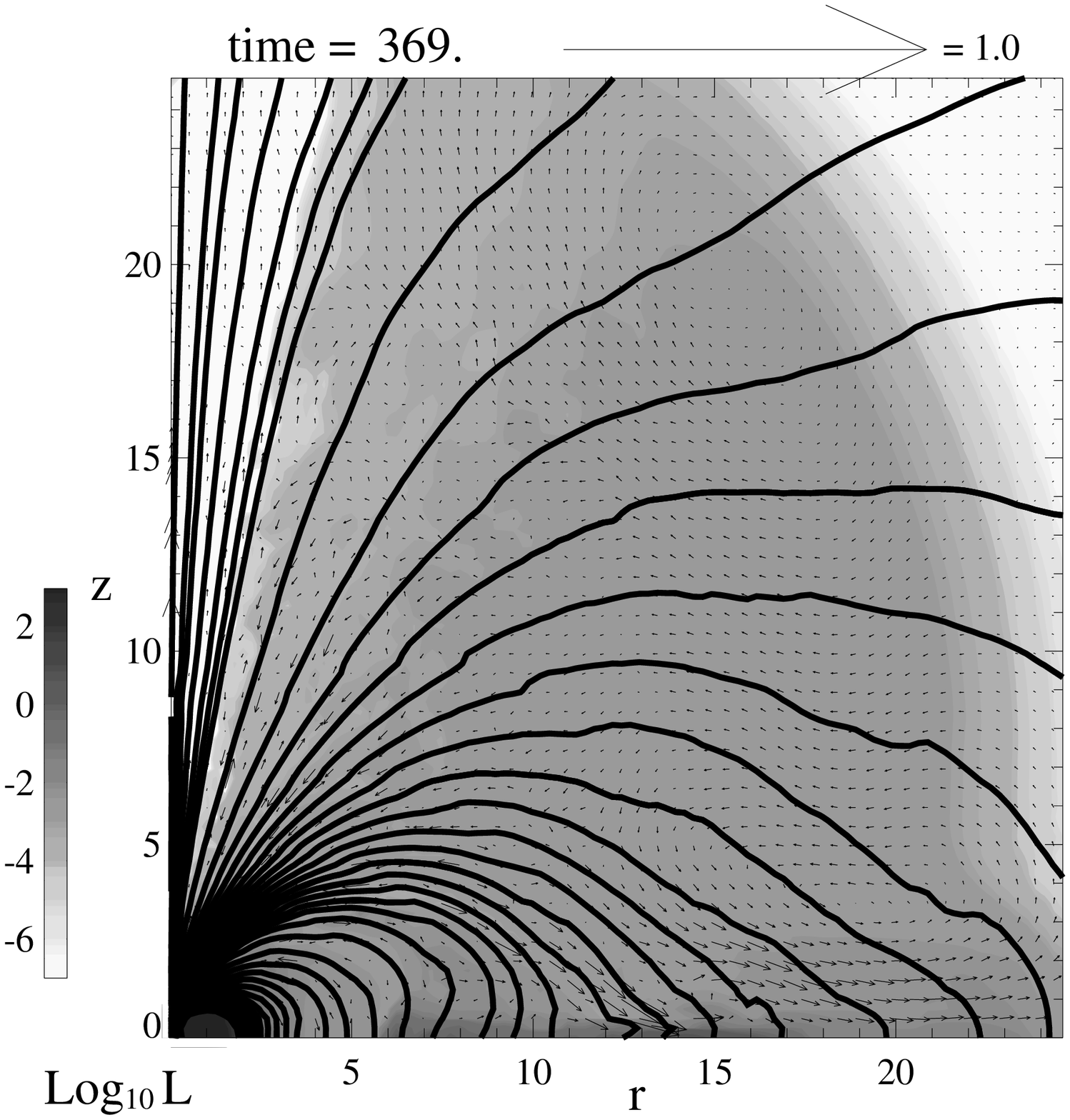}
\end{minipage}
  \caption{Time evolution of density distribution (upper
  panels) and angular momentum $L=\rho r v_{\varphi}$ distribution
  (lower panels) for Model-A.  The solid curves indicate the magnetic
  field lines.  The arrows show the velocity vectors.  Inside
  $r_{c}=13 r_{g}$, the magnetic torque removes angular momentum from
  the disk.  Outside $r_{c}$, the propeller action drives outflows.}
  \label{rotation:ps}
\end{minipage}
\end{figure}

Model-A includes the stellar rotation.  The rotation period is
parameterized by the corotation radius $r_{c}$.  The corotation radius
is taken to be $r_{c}=13 r_{g}$, which corresponds to 3 milliseconds
rotation period of the neutron star.  Figure \ref{rotation:ps} shows
the evolution of density and angular momentum distribution.  Outside
$r_{c}$, propeller action drives outflows. The magnetic field lines
deform its configurations into a shape similar to that of
helmet-streamer in the solar corona.  Inside $r_{c}$, the disk
material accretes because the magnetic field removes angular momentum
from the disk.  The deformed magnetic field lines around the
disk-magnetosphere boundary is similar to those of Ghosh-Lamb
model (Ghosh P., F. K. Lamb 1978).

\subsection*{Model-B : Slow-Rotator}

We neglect the stellar rotation by assuming that corotation radius is
far from the neutron star.  Figure \ref{norotation:ps} shows the
evolution of density and angular momentum distribution.  A consequence
of neglecting the stellar rotation is that the magnetic field
efficiently removes the angular momentum from the disk and deposits it
to the neutron star.  Since the surface layers of the disk most
efficiently lose their angular momentum, they accrete in dynamical
time-scale.  When the strength of magnetic field is not strong enough
to keep the Alfv\'{e}n radius outside the marginally stable radius
($r_{ms}$), the inner edge of the disk reaches $r_{ms}$.  During this
stage, we found the following quasi-periodic oscillations: (1) The
magnetic loops, which connect the neutron star and the disk, are
twisted by the rotation of the disk and expand.  (2) The expanding
loops prevent further accretion toward the neutron star and the
magnetosphere is loaded with the accreted material.  (3) When magnetic
twist exceeds critical angle, magnetic reconnection is triggered in
the current sheet formed inside the loops.  Magnetic reconnection
channels the accretion flow toward the neutron star, and the
magnetosphere is unloaded.  (4) This process repeats.

\begin{figure}[t]
p\begin{minipage}{180mm}
\begin{minipage}{50mm}
  \includegraphics[width=50mm]{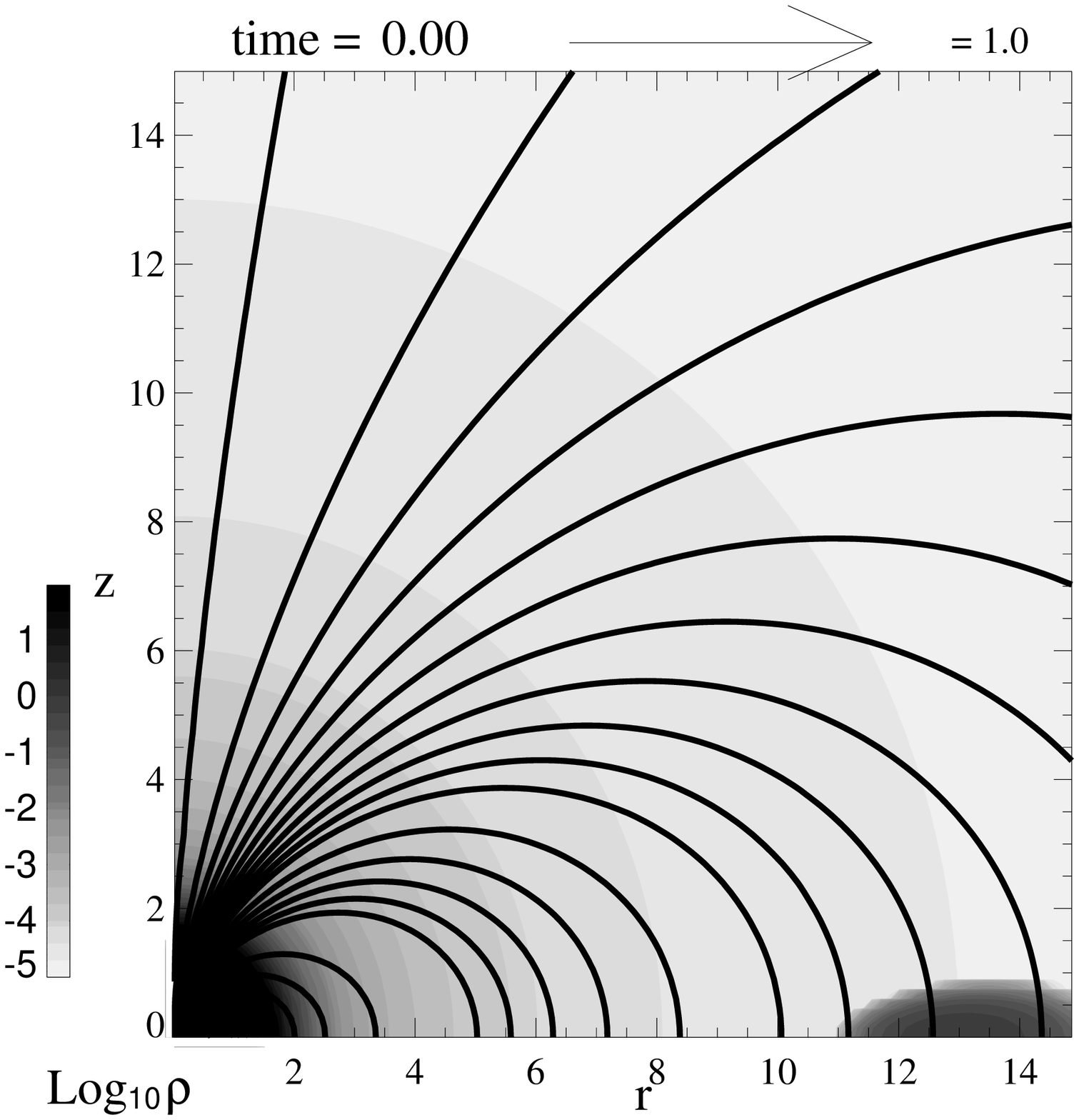}
\end{minipage}
\hfil\hspace{\fill}
\begin{minipage}{50mm}
  \includegraphics[width=50mm]{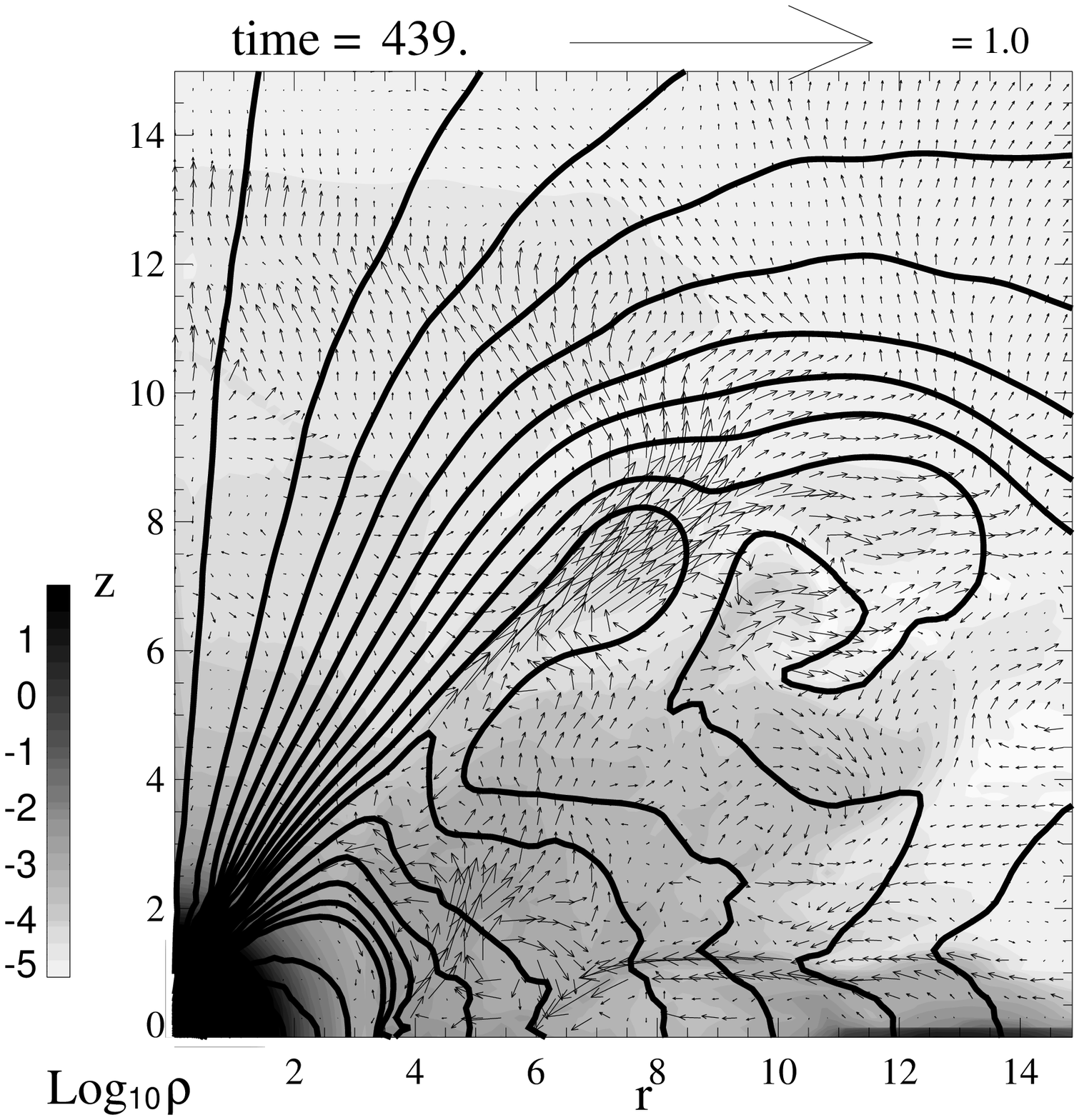}
\end{minipage}
\hfil\hspace{\fill}
\begin{minipage}{50mm}
  \includegraphics[width=50mm]{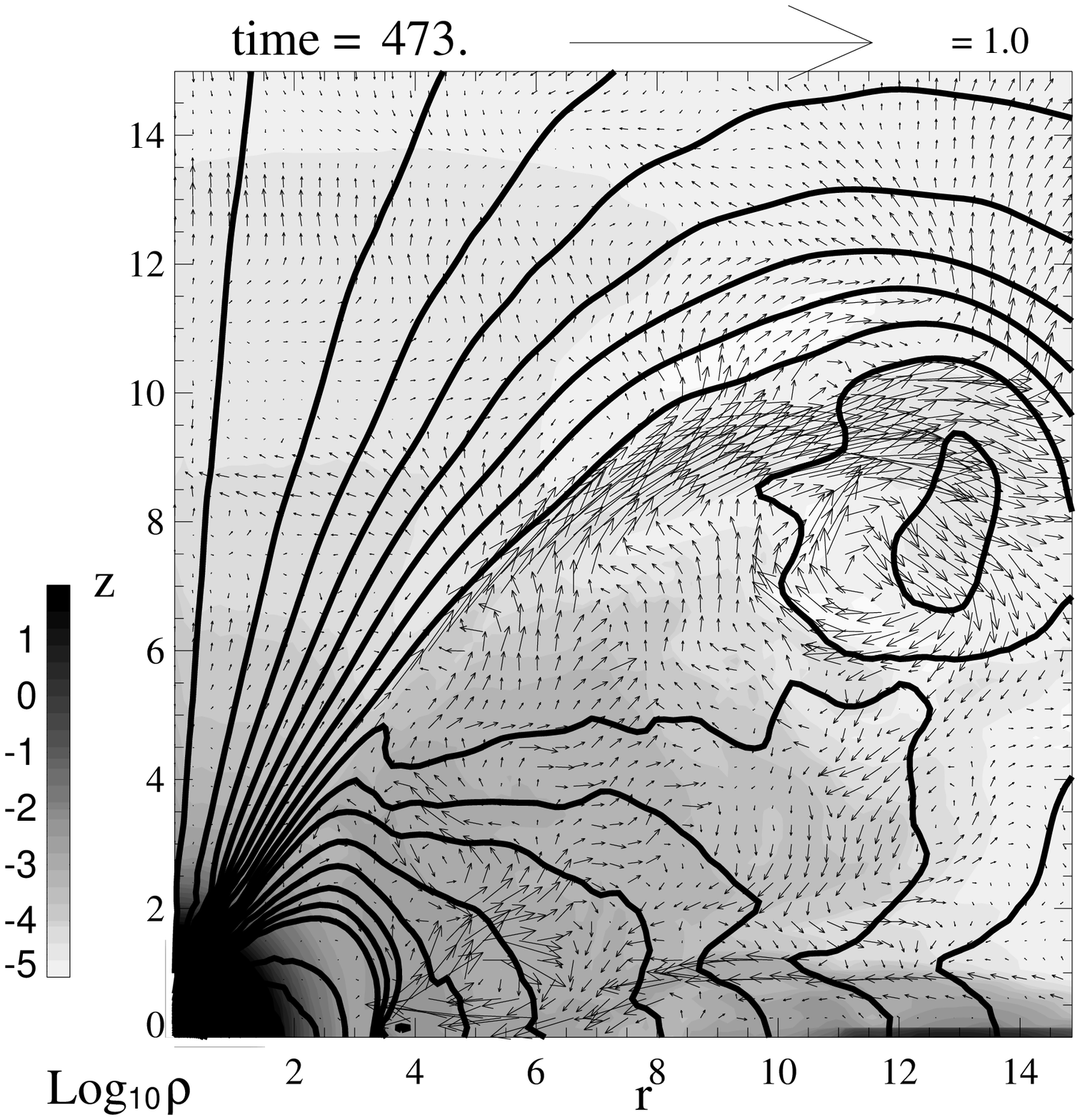}
\end{minipage}
\hfil\hspace{\fill}
\begin{minipage}{50mm}
  \includegraphics[width=50mm]{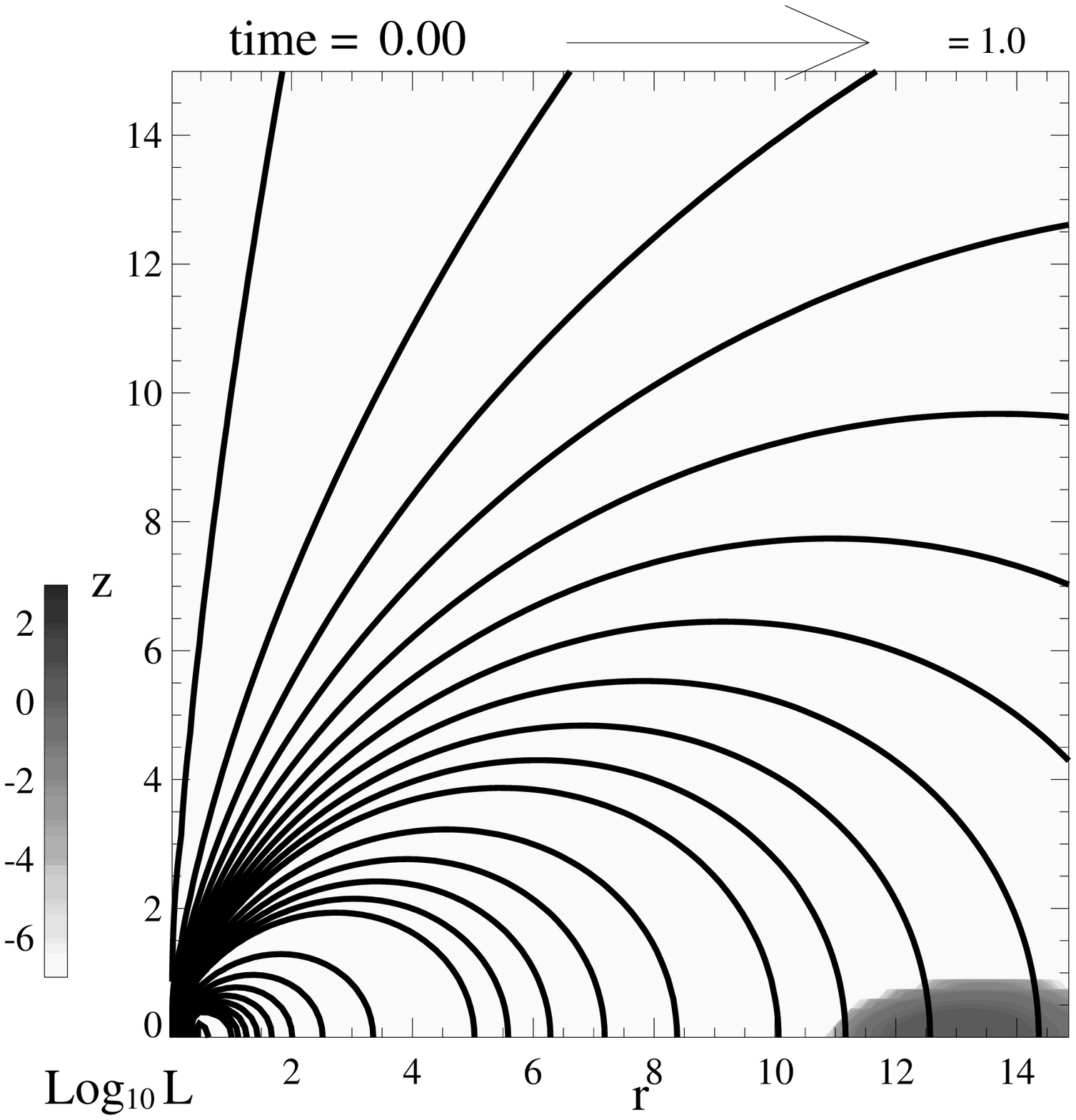}
\end{minipage}
\hfil\hspace{\fill}
\begin{minipage}{50mm}
  \includegraphics[width=50mm]{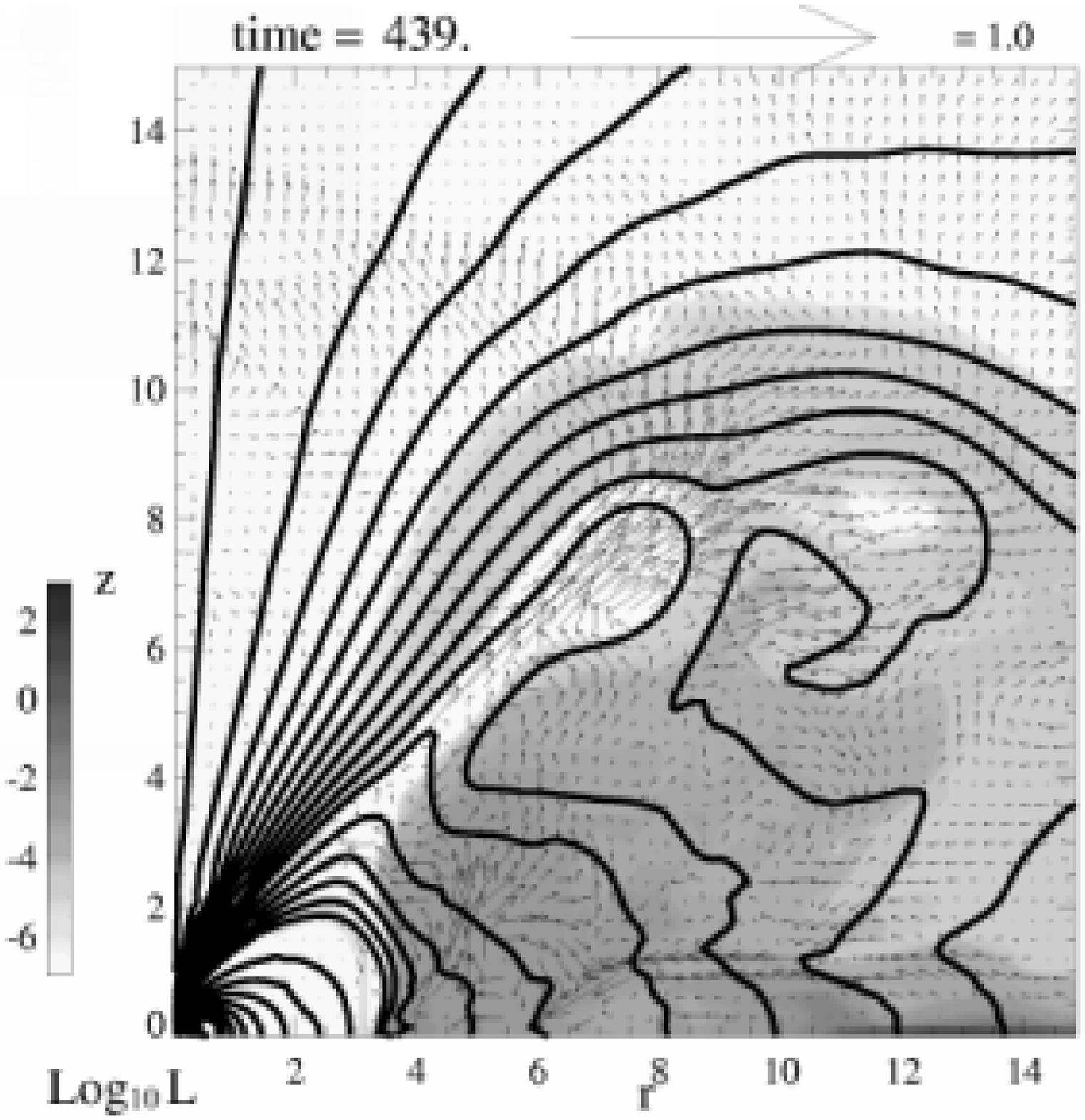}
\end{minipage}
\hfil\hspace{\fill}
\begin{minipage}{50mm}
  \includegraphics[width=50mm]{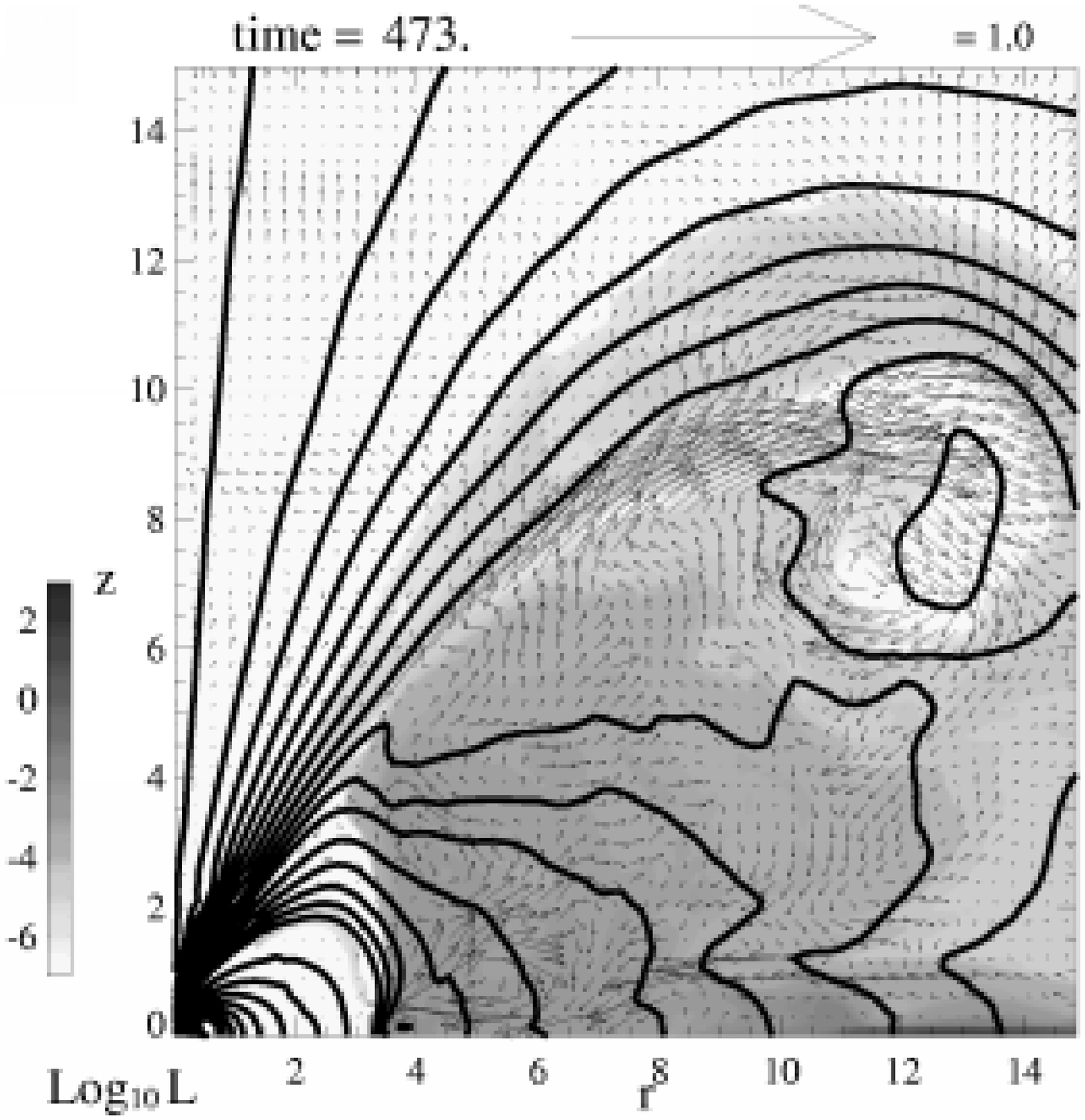}
\end{minipage}
\label{norotation:ps}
  \caption{Time evolution of density (upper panels) \& angular
  momentum (lower panels) distribution for Model-B.  The middle panels
  show the early stage of magnetic reconnection event and right panels
  show the ejection of the plasmoid after the reconnection.}
  \label{norotation:ps}
\end{minipage}
\end{figure}

\subsection*{Timing Analysis}

\begin{figure}[h]
\begin{minipage}{180mm}
  \includegraphics[width=180mm]{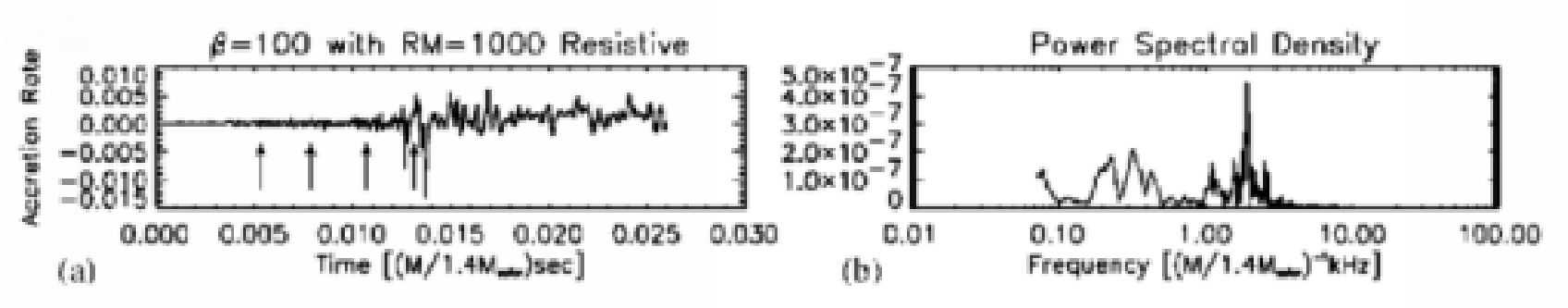}
  \label{psd:ps}
\end{minipage}
\caption{(a) Time variation of accretion rate for model-B.  Simulation
time $t_{sim}$ is converted to real time by $t=1.4\times
10^{-5}(M/1.4M_{\odot})t_{sim}$ sec.  Arrows indicate visible magnetic
reconnection events. (b) Power Spectral Density (PSD) of accretion
rate at (r, z)=(3$r_{g}$, 0).}
\label{psd:ps}
\vspace{5mm}
\begin{minipage}{180mm}
\includegraphics[width=180mm]{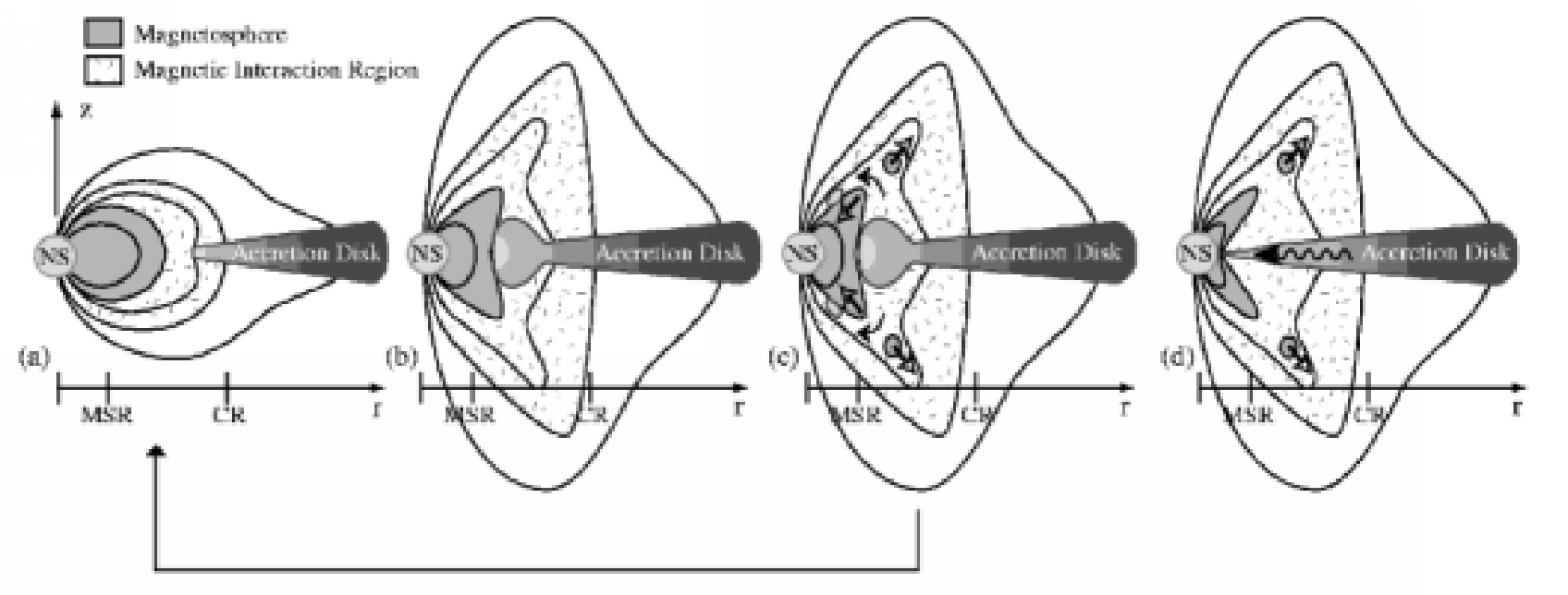}
  \caption{Mass accretion scenario with disk-magnetosphere
  interaction: CR and MSR stand for the corotation radius and
  marginally stable radius, respectively. (a) Inside the corotation
  radius, the disk material loses angular momentum and accretes.  (b)
  The accreted matter accumulates around the boundary between the
  magnetosphere and the disk.  The magnetic loops connecting the
  neutron star and the disk are twisted by the rotating plasma.  (c)
  When the magnetic twist exceeds critical angle, magnetic
  reconnection channels the accretion flow onto the magnetic poles of
  the neutron star and the magnetosphere is unloaded via
  accretion. The process (a) - (c) repeats.  (d) When the
  magnetosphere is collapsed, the radial disk oscillation modulates
  the accretion flow.}
  \label{scenario:ps}
\end{minipage}
\end{figure}

We analyzed the time variation of accretion rate falling into the
neutron star.  The left pannel of Figure \ref{psd:ps} shows the time
variation of the equatorial accretion rate at marginally stable radius
for Model-B.  The right pannel shows the power spectral density of the
time variation.  Typical frequencies are between 100 Hz and 2 kHz.
The arrows in the left pannel indicate the visible magnetic
reconnection events.

Our numerical results indicate that until the inner edge of the disk
reaches the marginally stable radius, the magnetospheric oscillation
which accompanies magnetic reconnection produces 100 - 1000 Hz QPOs.
Once the inner edge reaches the last stable orbit, magnetic
reconnection ceases.  In the latter stage, disk oscillation around 3
$r_{g}$ to 4 $r_{g}$ produces kHz QPOs (e.g., Matsumoto, R., S. Kato,
and F. Honma 1989).

\section*{SUMMARY}

Figure \ref{scenario:ps} summarizes the numerical results.  Inside the
corotation radius, magnetic torque removes angular momentum from the
disk and accretion proceeds.  The magnetic loops connecting the
neutron star and the disk expand and prevent the plasma from
further infall.  When the magnetic loops are twisted more than
critical angle, magnetic reconnection takes place inside the loops and
channels the accretion flow toward the neutron star.  After the
magnetosphere is unloaded via accretion, the process repeats (see
Goodson, A. P. \& R. M. Winglee 1999 for similar mechanism in
protostars).  In addition, this mechanism excites the radial disk
oscillation.  As a result, accretion rate changes quasi-periodically.
The typical frequency of this QPO is the Keplerian rotation period at
the inner edge of the disk, which corresponds to 100 - 1000 Hz when
the inner edge is located between 3 $r_{g}$ and 20 $r_{g}$.  When the
disk inner edge reaches the last stable orbit, the oscillation
frequency of the disk saturates around 1 - 2 kHz.  This scenario can
explain both the positive correlation of X-ray count rate and QPO
frequency and kHz QPOs observed by RXTE.

By extending our simulations to 3-D, we will be able to reproduce the
two peaks of QPOs conventionally explained by beat-frequency
interpretation.  Furthermore, we predict that QPO sources inevitably
accompany X-ray flares by magnetic reconnection and bipolar outflows
of hot X-ray emitting plasma similar to the optical jets in
protostars.

\section*{ACKNOWLEDGMENTS}
Numerical computations were carried out on VPP300/16R and VX/4R at the
Astronomical Data Analysis Center of the National Astronomical
Observatory, Japan.  This work is supported by Japan Science and
Technology Corporation (ACT-JST).

\end{document}